\documentclass[prd,showpacs,floatfix,twocolumn,,amsmath,amssymb,floatfix]{revtex4}
\usepackage{graphicx,color,dcolumn,booktabs,bm}
\usepackage{longtable,lscape}
\usepackage{txfonts}
\usepackage{overpic}
\usepackage{amssymb}
\usepackage{indentfirst}
\usepackage{epsfig}
\usepackage{feynmf}   
\usepackage{epstopdf}   
\usepackage{slashed}  
\usepackage{cases}
\usepackage{color}
\usepackage{multirow}
\usepackage{graphicx,color,dcolumn,booktabs,bm}

\graphicspath{{Figures/}} %

\graphicspath{{Figures/}} %
\usepackage[colorlinks, citecolor=blue,anchorcolor=red,menucolor=red, linkcolor=red,filecolor=red,runcolor=red,urlcolor=blue,frenchlinks=red]{hyperref}
\allowdisplaybreaks
\begin{document}

\title{The masses and axial currents of the doubly charmed baryons}

\author{Zhi-Feng Sun$^{1,2}$}\email{sunzhif09@lzu.edu.cn}
\author{Zhan-Wei Liu$^{3}$}\email{zhan-wei.liu@adelaide.edu.au}
\author{Xiang Liu$^{1,2}$}
\email{xiangliu@lzu.edu.cn}
\author{Shi-Lin Zhu$^{4,5,6}$}
\email{zhusl@pku.edu.cn} \affiliation{
$^1$School of Physical Science and Technology, Lanzhou University, Lanzhou 730000, China\\
$^2$Research Center for Hadron and CSR Physics, Lanzhou University and Institute of Modern Physics of CAS, Lanzhou 730000, China\\
$^3$CSSM, School of Chemistry and Physics, University of Adelaide, Adelaide, South Australia 5005, Australia
$^4$School of Physics and State Key Laboratory of Nuclear Physics and Technology, Peking University, Beijing 100871, China \\
$^5$Collaborative Innovation Center of Quantum Matter, Beijing 100871, China \\
$^6$Center of High Energy Physics, Peking University, Beijing
100871, China }

\begin{abstract}
The chiral dynamics of the doubly heavy baryons is solely governed
by the light quark. In this work, we have derived the chiral
corrections to the mass of the doubly heavy baryons up to N$^3$LO.
The mass splitting of $\Xi_{cc}$ and $\Omega_{cc}$ at the N$^2$LO
depends on one unknown low energy constant $c_7$.
{By fitting the lattice masses of $\Xi_{cc}(3520)$, we estimate the
mass of $\Omega_{cc}$ to be around 3.726 GeV.}
Moreover, we have also
performed a systematical analysis of the chiral corrections to the
axial currents and axial charges of the doubly heavy baryons. The
chiral structure and analytical expressions will be very useful to
the chiral extrapolations of the future lattice QCD simulations of
the doubly heavy baryons.
\end{abstract}


\maketitle

\section{Introduction}\label{sec1}

As one of the most important groups in the baryon family, the doubly
charmed baryons are composed of two charmed quarks and one light
quark (the doubly heavy baryons $\Xi_{cc}^{++}$, $\Xi_{cc}^{+}$ and
$\Omega_{cc}^{+}$ with quark components $ccu$, $ccd$ and $ccs$,
respectively), which were predicted in the quark model (see Ref.
\cite{Moinester:1995fk} for a detailed review). In the past decades,
there have been some experimental efforts in the search of the
doubly charmed baryons
\cite{Mattson:2002vu,Moinester:2002uw,Ocherashvili:2004hi,Engelfried:2005kd}.
The SELEX Collaboration announced the first observation of the
doubly charmed baryon $\Xi_{cc}^+(3520)$ with the mass $M=3519\pm1$
MeV and width $\Gamma=3$ MeV \cite{Mattson:2002vu}, where the
observed decay mode is  $\Xi_{cc}^+\to \Lambda_c^+K^-\pi^+$. Later,
$\Xi_{cc}^+(3520)$ was confirmed by SELEX in the $pD^+K^-$ decay
channel with the mass $3518.7\pm1.7$ MeV \cite{Ocherashvili:2004hi}.
Although SELEX also reported $\Xi_{cc}^+(3520)$, these results were
not confirmed by FOCUS \cite{Ratti:2003ez}, BaBar
\cite{Aubert:2006qw}, Belle \cite{Chistov:2006zj} and LHCb
collaborations \cite{Aaij:2013voa}.

The doubly charmed baryons have been extensively studied with
different theoretical approaches. The $\Xi_{cc}$ mass was predicted
to be $3.48\sim 3.74$ GeV in the quark model, while the
$\Omega_{cc}$ mass is estimated to be $3.59\sim 3.86$ GeV
\cite{Roncaglia:1995az, Ebert:1996ec, SilvestreBrac:1996wp,
Tong:1999qs, Gerasyuta:1999pc, Itoh:2000um, Kiselev:2001fw,
Narodetskii:2002ib, Ebert:2002ig, Vijande:2004at, Migura:2006ep,
Albertus:2006ya, Roberts:2007ni}. The Lattice QCD groups also
studied these systems \cite{Lewis:2001iz, Na:2008hz, Liu:2009jc,
Namekawa:2012mp, Alexandrou:2012xk}, where the predicted mass of
$\Xi_{cc}$ is $3.51\sim 3.67$ GeV and the mass of $\Omega_{cc}$ is
$3.68\sim 3.76$ GeV.

The mass splittings of baryons within the same multiplet encode
important information on their inner structure. For example, the
mass splittings of the light baryons were reviewed in Refs.
\cite{Scherer:2002tk,Bernard:1995dp}. In Refs. \cite{Guo:2008ns,
Jiang:2014ena}, the mass splitting of the singly heavy baryons was
studied within the framework of the chiral perturbation theory. In
Ref. \cite{Brodsky:2011zs}, the authors investigated the mass
splitting of the doubly heavy baryons by considering the heavy
diquark symmetry. Besides the baryon mass, the axial current and
axial charge of the baryons are also very important observables,
which attract lots of attention
\cite{Jiang:2008aqa,Jiang:2008we,Aoki:2009ri,Jiang:2009sf,Aoki:2009kt,Detmold:2011rb,Detmold:2012ge,FloresMendieta:2012dn,FloresMendieta:2000mz,Zhu:2000zf,Beane:2003xv,
Beane:2004rf,Tiburzi:2005hg,FloresMendieta:2006ei,Schindler:2006it,Park:1993jf,Luty:1993gi,Borasoy:1998pe,Oh:1999yj,Zhu:2002tn,Bijnens:1985kj,Jenkins:1990jv,Jenkins:1991es}
.

The experimental search of the doubly charmed baryons is full of
challenges and opportunities. In this work, we adopt the chiral
perturbation theory to calculate the chiral corrections to the
doubly charmed baryon masses and their mass splittings, which will
be helpful to further experimental exploration of the doubly charmed
baryons. Under the same framework, we also study the chiral
corrections to the axial charge and axial current of the doubly
charmed baryons, which may be measured through the semileptonic
decays of the doubly charmed baryons in the future.

Chiral perturbation theory ($\chi$PT) is an elegant framework to
deal with the low energy process in hadron physics. With the help of
the chiral power counting scheme proposed by Weinberg {\it et al}.
\cite{Weinberg:1991um, Ecker:1994gg}, one can consider the chiral
corrections to the physical observables order by order.

In the baryon sector, the baryon mass does not vanish in the chiral
limit. This inherent mass scale breaks the naive chiral power
counting. To solve this issue, various schemes were proposed such as
the heavy baryon $\chi$PT, infrared baryon $\chi$PT, and extended
on-mass-shell method etc.

In the heavy baryon $\chi$PT, the baryon is treated to be extremely
heavy and acts as a static source \cite{Jenkins:1990jv}, which
allows us to take the non-relativistic limit of the fully
relativistic theory and make expansion in powers of the inverse
baryon mass. For the case of infrared regularization, the loop
integral can be separated into infrared regular part and infrared
singular one \cite{Becher:1999he, Kubis:2000zd}, where the later one
conserves the Weinberg's power counting rule. In the extended
on-mass-shell method, the power counting breaking terms are
subtracted and the low energy constants are redefined
\cite{Geng:2008mf, MartinCamalich:2010fp, Geng:2010df, Geng:2010yc}.
In our work, we use the heavy baryon $\chi$PT approach to
investigate the chiral corrections to the masses and axial currents
of the doubly charmed baryons.

This paper is organized as follows. After the introduction, we
introduce the chiral Lagrangians of the doubly charmed baryons and
its non-relativistic reduction in Sec. II\label{sec2}. Then we present
the calculation details of the chiral corrections to the masses and
axial currents of the doubly charmed baryons and the corresponding
numerical results in Secs. III\label{sec3} and IV\label{sec4}
respectively. This work ends with a summary in Sec. V\label{sec5}. We
collect the N$^3$LO chiral corrections to the masses and some
lengthy expressions in the appendix.

\section{The chiral Lagrangians of the doubly
charmed baryons\label{sec2}}

In order to calculate the chiral corrections to the masses and axial
currents , we need construct the chiral effective Lagrangians of the
doubly charmed baryons with the help of chiral, parity and charge
conjugation symmetries. We firstly introduce the notations $U$ and
$u$ to describe the psudoscalar meson field, which have the relation
\begin{eqnarray}
U=u^2=\exp\left(i\frac{\phi(x)}{F_0}\right),
\end{eqnarray}
where $\phi(x)$ has the definition
\begin{eqnarray}
\phi(x)=\left(
          \begin{array}{ccc}
            \pi^0+\frac{1}{\sqrt{3}}\eta & \sqrt{2}\pi^+ & \sqrt{2}K^+ \\
            \sqrt{2}\pi^- & -\pi^0+\frac{1}{\sqrt{3}}\eta & \sqrt{2}K^0 \\
            \sqrt{2}K^- & \sqrt{2}\bar{K^0} & -\frac{2}{\sqrt{3}}\eta \\
          \end{array}
        \right).
\end{eqnarray}
The doubly heavy baryon field $\psi$ with spin $\frac{1}{2}$ is a
column vector in the flavor space, i.e.
\begin{eqnarray}
\psi &=&\left(
          \begin{array}{c}
            \Xi_{cc}^{++} \\
            \Xi_{cc}^{+} \\
            \Omega_{cc}^+ \\
          \end{array}
        \right),
\end{eqnarray}
where the quark contents of $\Xi_{cc}^{++}$, $\Xi_{cc}^+$, and
$\Omega_{cc}^+$ are $ccu$, $ccd$, and $ccs$, respectively.

\renewcommand{\arraystretch}{1.4}
\begin{table}[htbp]
\caption{The properties of the building blocks under the
$SU(3)_L\times SU(3)_R$ (CH), parity (P) and charge conjugation (C)
transformations.\label{tab1}}
\begin{tabular}{ccccccc}
\toprule[1pt]

  &$U$           &$u$         &$\chi$            &$f^{R}_{\mu\nu}$           &$f^{L}_{\mu\nu}$&$D_\mu \psi$           \\\midrule[0.5pt]
CH&$V_RUV_L^\dag$&$V_RuK^\dag$&$V_R\chi V_L^\dag$&$V_Rf^{R}_{\mu\nu}V_R^\dag$&$V_Lf^{L}_{\mu\nu}V_L^\dag$&$KD_\mu \psi$\\
P &$U^\dag$      &$u^\dag$    &$\chi^\dag$       &$f^{L\mu\nu}$&$f^{R\mu\nu}$&$\gamma^0D^\mu \psi$\\
C &$U^T$
&$u^T$&$\chi^T$&$-(f^{L}_{\mu\nu})^T$&$-(f^{R}_{\mu\nu})^T$&$CD^{\prime
T}_\mu \bar{\psi}^T$
\\\bottomrule[1pt]
  &$\psi $&$\bar{\psi}$      &$\chi_{\pm}$       &$f^{\pm}_{\mu\nu}$
  &$u_\mu$&$\Gamma_\mu$\\\midrule[0.5pt]
CH&$K\psi$&$\bar{\psi}K^\dag$&$K\chi_{\pm}K^\dag$&$Kf^{\pm}_{\mu\nu}K^\dag$&$Ku_\mu K^\dag$&$K\Gamma^\mu K^\dag-\partial^\mu KK^\dag$\\
P &$\gamma^0\psi$&$\bar{\psi}\gamma^0$&$\pm \chi_{\pm}$&$\pm f^{\pm \mu\nu}$&$-u^\mu$&$\Gamma^\mu$\\
C &$C\bar{\psi}^T$&$\psi^TC$&$\chi_{\pm}^T$&$\mp
(f^{\pm}_{\mu\nu})$&$(u_\mu)^T$&$-(\Gamma_\mu)^T$\\\bottomrule[1pt]
\end{tabular}
\end{table}

In Table \ref{tab1}, we show the transformation properties of the
building blocks, which include $\chi, \chi_{\pm}, f^R_{\mu\nu},
f^L_{\mu\nu}, f^\pm_{\mu\nu}, u_\mu, \Gamma_\mu, D_\mu$ and
$D^\prime_\mu$ with the definitions
\begin{eqnarray}
\chi&=&2B_0(s+ip),\\
\chi_{\pm}&=&u^\dag\chi u^\dag\pm u\chi^\dag u,\\
f^R_{\mu\nu}&=&\partial_\mu r_\nu-\partial_\nu r_\mu-i[r_\mu,r_\nu],\\
f^L_{\mu\nu}&=&\partial_\mu l_\nu-\partial_\nu l_\mu-i[l_\mu,l_\nu],\\
f^\pm_{\mu\nu}&=&u^\dag f^R_{\mu\nu} u\pm uf^L_{\mu\nu} u^\dag,\\
u_\mu&=&i[u^\dag(\partial_\mu-ir_\mu)u-u(\partial_\mu-il_u)u^\dag],\\
\Gamma_\mu&=&\frac{1}{2}[u^\dag(\partial_\mu-ir_\mu)u+u(\partial_\mu-il_u)u^\dag],\\
D_\mu&=&\partial_\mu+\Gamma_\mu-iv^{(s)}_\mu,\\
D^\prime_\mu&=&\partial_\mu-\Gamma_\mu+iv^{(s)}_\mu,
\end{eqnarray}
where $r_\mu=v_\mu+a_\mu$, $l_\mu=v_\mu-a_\mu$, and $v_\mu,
v^{(s)}_\mu, a_\mu, s, p$ are external $c$-number fields.
Considering the transformation properties listed in Table
\ref{tab1}, the chiral Lagrangian of the doubly heavy baryon can be
constructed order by order, i.e.,
\begin{eqnarray}
\mathcal{L}^{(1)}&=&\bar{\psi}(i{D\!\!\!\slash}-m+\frac{g_A}{2}\gamma^\mu\gamma_5u_\mu)\psi,\label{l1}\\
\nonumber \mathcal{L}^{(2)}&=&c_1\bar{\psi}\langle
\chi_\pm\rangle\psi-\left\{\frac{c_2}{8m^2} \bar{\psi}\langle u_\mu
u_\nu \rangle \{ D^\mu,D^\nu\}\psi+h.c.\right\}\\\nonumber
&&-\left\{\frac{c_3}{8m^2} \bar{\psi}\{ u_\mu, u_\nu\}  \{
D^\mu,D^\nu\}\psi+h.c.\right\}+\frac{c_4}{2}\bar{\psi}\langle
u^2\rangle\psi\\\nonumber &&+\frac{c_5}{2}\bar{\psi} u^2\psi+\left\{
\frac{ic_6}{4}\bar{\psi}\sigma^{\mu\nu}[u_\mu,u_\nu]\psi+h.c.\right\}+c_7\bar{\psi}\hat{\chi_+}\psi\\
&&+\frac{c_8}{8m}\bar{\psi}\sigma^{\mu\nu}f^+_{\mu\nu}\psi+\frac{c_9}{8m}\bar{\psi}\sigma^{\mu\nu}\langle
f^+_{\mu\nu}\rangle\psi\label{l2}
\\\nonumber
\mathcal{L}^{(3)}&=&\bar{\psi}\left\{\frac{h_1}{2}\gamma^\mu
\gamma_5\langle \chi_+\rangle u_\mu+\frac{h_2}{2}\gamma^\mu
\gamma_5\{\hat{\chi}_+,u_\mu\}+\frac{h_3}{2}\gamma^\mu\gamma_5\langle
\hat{\chi}_+u_\mu\rangle\right.\\
&&\left. +...\right\}\psi,
\\\nonumber
\mathcal{L}^{(4)}&=&e_1\bar{\psi}\langle \chi_+\rangle\langle
\chi_+\rangle\psi+e_2\bar{\psi} \hat{\chi}_+\langle
\chi_+\rangle\psi+e_3\bar{\psi}\langle \hat{\chi}_+
\hat{\chi}_+\rangle\psi\\\nonumber &&+e_4\bar{\psi} \hat{\chi}_+
\hat{\chi}_+\psi+e_5\bar{\psi}\langle \chi_-\rangle\langle
\chi_-\rangle\psi+e_6\bar{\psi} \hat{\chi}_-\langle
\chi_-\rangle\psi\\ &&+e_7\bar{\psi}\langle \hat{\chi}_-
\hat{\chi}_-\rangle\psi+e_8\bar{\psi} \hat{\chi}_-
\hat{\chi}_-\psi+....
\end{eqnarray}
where $\hat{A}=A-\frac{1}{3}\langle A\rangle$ and $\langle A\rangle$
denotes the trace of $A$ in the flavor space. In the above
Lagrangians, $c_i$ ($i=1,\cdots,9$), $h_j$ ($j=1,\cdots, 3$), and
$e_k$ ($k=1,\cdots, 8$) are the effective coupling constants. They
are sometimes denoted as the low energy constants (LECs).

Since the doubly heavy baryons are very heavy, we can take the
non-relativistic limit of the fully relativistic theory and expands
the Lagrangian in power of the inverse of the doubly heavy baryon
mass. The four-momentum of the doubly heavy baryon can be written as
\begin{equation}
p_\mu=mv_\mu+l_\mu
\end{equation}
where $v_\mu$ is the four-velocity and $l_\mu$ the small off-shell
momentum, which satisfies $v\cdot l\ll m$. The baryon field is
decomposed into the light and heavy components $\psi=e^{-imv\cdot
x}(H+h)$, where $v\!\!\!\slash H=H$, $ v\!\!\!\slash h=-h$.

The generating functional for the relativistic theory reads
\begin{eqnarray}
\exp i Z[\eta,\bar{\eta},v,a,s,p]&=&\int
[d\psi][d\bar{\psi}][du]\exp
\bigg\{i\Big[S\nonumber\\
&&+\int d^4x(\bar{\eta}\psi+\bar{\psi}\eta)\Big]\bigg\},
\end{eqnarray}
where
\begin{eqnarray}
S&=&\int d^4x \mathcal{L}.
\end{eqnarray}
In the terms of the fields $H$ and $h$, we can rewrite the original
Lagrangian
\begin{eqnarray}
\mathcal{L}&=&\bar{H}\mathcal{A}H+\bar{h}\mathcal{B}H+\bar{H}\gamma^0\mathcal{B}^\dag\gamma^0h-\bar{h}\mathcal{C}h.\label{hq}
\end{eqnarray}
$\mathcal{A}$, $\mathcal{B}$ and $\mathcal{C}$ in Eq. (\ref{hq}) can
be expanded in series of terms of different orders of $q^i$, where
$q$ is the low energy momentum,
\begin{eqnarray}
\mathcal{A}&=&\mathcal{A}_{(1)}+\mathcal{A}_{(2)}+...,\label{t1}\\
\mathcal{B}&=&\mathcal{B}_{(1)}+\mathcal{B}_{(2)}+...,\\
\mathcal{C}&=&\mathcal{C}_{(1)}+\mathcal{C}_{(2)}+....\label{t3}
\end{eqnarray}
The expressions of $\mathcal{A}, \mathcal{B}, \mathcal{C}$ are
collected in Appendix. With the replacement
\begin{eqnarray}
R&=&\frac{1}{2}(1+v\!\!\!\slash)e^{imv\cdot x}\eta,\nonumber\\
\rho&=&\frac{1}{2}(1-v\!\!\!\slash)e^{imv\cdot x}\eta,\nonumber
\end{eqnarray}
we have
\begin{eqnarray}
\bar{\eta}\psi +\bar{\psi}\eta
&=&\bar{R}H+\bar{H}R+\bar{\rho}h+\bar{h}\rho.
\end{eqnarray}
With $h^{\prime}=h-\mathcal{C}^{-1}(\mathcal{B}H+\rho)$ and after
integrating out the heavy degrees of freedom, the generating
functional becomes
\begin{eqnarray}
\exp i Z[R, \bar{R}, \rho, \bar{\rho}, v, a, s, p]&=&\int
[dH][d\bar{H}][du]\Delta_h\exp i\bigg [ S^\prime\nonumber\\
&&+\int d^4x(\bar{R}H+\bar{H}R)\bigg],
\end{eqnarray}
where
\begin{eqnarray}
S^\prime &=&\int d^4x
\bar{H}\Big[\mathcal{A}+(\gamma_0\mathcal{B}^\dag\gamma_0)\mathcal{C}^{-1}\mathcal{B}\Big]H
\end{eqnarray}
and $\Delta_h$ is a constant. Then, one expands the matrix
$\mathcal{C}^{-1}$ in terms of $1/m$
\begin{eqnarray}
\mathcal{C}^{-1}&=&\frac{1}{2m}-\frac{i(v\cdot D)+g_AS_v\cdot
u}{(2m)^2}-\frac{\mathcal{C}_{(2)}}{(2m)^2}\nonumber\\
&&+\frac{(iv\cdot D+g_AS_v\cdot u)^2}{(2m)^3} + \cdots.
\end{eqnarray}
Finally, the non-relativistic Lagrangians corresponding to the
action $S^\prime$ can be expressed as
\begin{eqnarray}
\mathcal{L}^\prime &=&\mathcal{L}^\prime _{(1)}+\mathcal{L}^\prime
_{(2)}+\mathcal{L}^\prime _{(3)}+\mathcal{L}^\prime _{(4)}+\cdots
\label{m1}
\end{eqnarray}
with $\mathcal{L}^\prime _{(i)}=\bar{H}T_{(i)}H$
($i=1,2,3,4,\cdots$), where
\begin{eqnarray}
T_{(1)} &=&i(v\cdot D)+g_AS_v\cdot u,
\end{eqnarray}
\begin{eqnarray}
T_{(2)} &=&c_1\langle \chi_+\rangle+\frac{c_2}{2}\langle (v\cdot
u)^2\rangle+c_3(v\cdot u)^2+\frac{c_4}{2}\langle
u^2\rangle\nonumber\\
&&+\frac{c_5}{2}u^2+\frac{c_6}{2}[S_v^\mu, S_v^\nu][u_\mu,
u_\nu]+c_7\hat{\chi_+}\nonumber\\
&&-\frac{ic_8}{4m}[S_v^\mu,
S_v^\nu]f^+_{\mu\nu}-\frac{ic_9}{4m}[S_v^\mu, S_v^\nu]\langle
f^+_{\mu\nu}\rangle\nonumber\\
&&+\frac{2}{m}(S_v\cdot D)^2-\frac{ig_A}{2m}\{S_v\cdot D, v\cdot
u\}\nonumber\\
&&-\frac{g_A^2}{8m}(v\cdot u)^2+...,
\end{eqnarray}
\begin{eqnarray}
T_{(3)}&=&h_1S_v^\mu \langle \chi_+ \rangle u_\mu+h_2S_v^\mu\{\hat{\chi}_+,u_\mu\}+h_3S_v^\mu\langle \hat{\chi}_+u_\mu\rangle\nonumber\\
&&-\frac{c_8}{4m^2}S_v\cdot D(v^\mu S_v^\nu-v^\nu
S_v^\mu)f^+_{\mu\nu}\nonumber\\
&&-\frac{c_9}{4m^2}S_v\cdot D(v^\mu S_v^\nu-v^\nu S_v^\mu)\langle
f^+_{\mu\nu}\rangle\nonumber\\
&&+\frac{ig_Ac_8}{16m^2}(v^\mu S_v^\nu-v^\nu S_v^\mu)[v\cdot
u,f^+_{\mu\nu}]\nonumber\\
&&\frac{c_8}{4m^2}(v^\mu S_v^\nu-v^\nu S_v^\mu)f^+_{\mu\nu}S_v\cdot
D\nonumber\\
&&+\frac{c_9}{4m^2}(v^\mu S_v^\nu-v^\nu S_v^\mu)\langle
f^+_{\mu\nu}\rangle S_v\cdot D\nonumber\\
&&-\frac{i}{m^2}S_v\cdot Dv\cdot DS_v\cdot D+...,
\end{eqnarray}
\begin{eqnarray}
T_{(4)}&=&e_1\langle \chi_+\rangle \langle \chi_+\rangle +e_2\hat{
\chi_+}\langle \chi_+\rangle+e_3\langle \hat{ \chi_+} \hat{
\chi_+}\rangle+e_4\hat{ \chi_+} \hat{ \chi_+}\nonumber\\
&&+e_5\langle \chi_-\rangle \langle \chi_-\rangle+e_6\hat{
\chi_-}\langle \chi_-\rangle+e_7\langle \hat{ \chi_-} \hat{
\chi_-}\rangle+e_8\hat{ \chi_-} \hat{ \chi_-}\nonumber\\
&&-\frac{c_8^2}{32m^3}(v^\mu S_v^\nu-v^\nu
S_v^\mu)f^+_{\mu\nu}(v^\alpha S_v^\beta-v^\beta
S_v^\alpha)f^+_{\alpha\beta}\nonumber\\
&&-\frac{c_8c_9}{32m^3}(v^\mu S_v^\nu-v^\nu
S_v^\mu)f^+_{\mu\nu}(v^\alpha S_v^\beta-v^\beta S_v^\alpha)\langle
f^+_{\alpha\beta}\rangle\nonumber\\
&&-\frac{c_8c_9}{32m^3}(v^\mu S_v^\nu-v^\nu S_v^\mu)\langle
f^+_{\mu\nu}\rangle(v^\alpha S_v^\beta-v^\beta
S_v^\alpha) f^+_{\alpha\beta}\nonumber\\
&&-\frac{c_9^2}{32m^3}(v^\mu S_v^\nu-v^\nu S_v^\mu)\langle
f^+_{\mu\nu}\rangle(v^\alpha S_v^\beta-v^\beta S_v^\alpha)\langle
f^+_{\alpha\beta}\rangle\nonumber\\
&&+\frac{ic_8}{(2m)^3}S_v\cdot Dv\cdot D(v^\mu S_v^\nu-v^\nu
S_v^\mu)f^+_{\mu\nu}\nonumber\\
&&+\frac{ic_9}{(2m)^3}S_v\cdot Dv\cdot D(v^\mu S_v^\nu-v^\nu
S_v^\mu)\langle f^+_{\mu\nu}\rangle\nonumber\\
&&-\frac{ic_8}{(2m)^3}(v^\mu S_v^\nu-v^\nu
S_v^\mu)f^+_{\mu\nu}v\cdot DS_v\cdot
D\nonumber\\
&&-\frac{ic_9}{(2m)^3}(v^\mu S_v^\nu-v^\nu
S_v^\mu)\langle f^+_{\mu\nu}\rangle v\cdot DS_v\cdot D\nonumber\\
&&+\frac{c_7}{m^2}S_v\cdot D\hat{\chi_+}S_v\cdot
D+\frac{c_1}{m^2}S_v\cdot D\langle\chi_+\rangle S_v\cdot D\nonumber\\
&&-\frac{ic_8}{4m^3}S_v\cdot D[S_v^\mu, S_v^\nu]f^+_{\mu\nu}S_v\cdot
D\nonumber\\
&&-\frac{ic_9}{4m^3}S_v\cdot D[S_v^\mu, S_v^\nu]\langle
f^+_{\mu\nu}\rangle S_v\cdot
D\nonumber\\
&&-\frac{1}{2m^3}S_v\cdot D(v\cdot D)^2S_v\cdot D+...,
\end{eqnarray}
where
\begin{equation}
S_v^\mu=\frac{i}{2}\gamma_5\sigma^{\mu\nu}v_\nu.
\end{equation}
The above Lagrangians will be employed to calculate the chiral
correction to the mass and axial current of the doubly charmed
baryons.

\section{The chiral correction to the mass of
the doubly heavy baryon}\label{sec3}

\begin{figure}[htpb]
\centering
\includegraphics[width=0.9\linewidth]{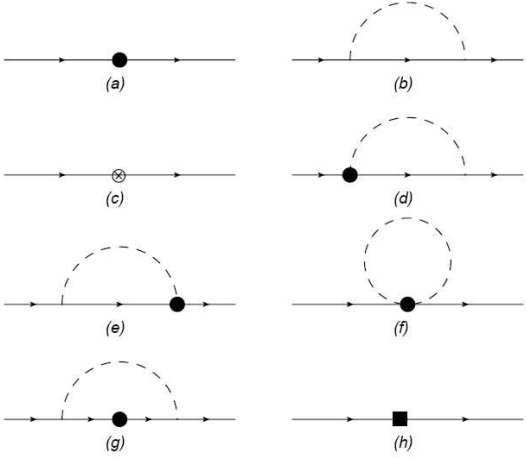}
\caption{The Feynman diagrams which contribute to the self-energy of
doubly charmed baryon. The solid and dashed lines denote the doubly
charmed baryons and goldstone bosons. The solid dot, circle-cross
and black box denote the vertices from the ${\cal O}(p^2, p^3, p^4)$
Lagrangians respectively. \label{fig1}}
\end{figure}

With the notations $\eta=v\cdot p$ and $\xi=(p-mv)^2$, the full
propagator of the doubly heavy baryon is written as
\begin{eqnarray}
G&=&\frac{i}{v\cdot p-m_0-\Sigma_B(\eta,\xi)}\nonumber\\
&=&\frac{iZ_N}{v\cdot
p-m-Z_N\tilde{\Sigma}_B(\eta,\xi)},\label{self}
\end{eqnarray}
where $\Sigma_B(\eta,\xi)$ denotes the high order contributions to
the self-energy, which are from Fig. \ref{fig1} (a-h).

The mass of the doubly charmed baryon is
\begin{equation}
m=m_0+\Sigma_B(0,0). \label{eqmSum}
\end{equation}
And the renormalization constant reads
\begin{equation}
Z_N=\frac{1}{1-\Sigma_B^\prime(0,0)}
\end{equation}
with
\begin{equation}
\Sigma_B^\prime(0,0)=\frac{\partial \Sigma_B(\eta,\xi)}{\partial
\eta}\big|_{(\eta,\xi)=(0,0)}.
\end{equation}
The chiral contribution to the self-energy up to
next-to-next-to-leading order (NNLO) includes three pieces
\begin{eqnarray}
\Sigma^{(a)}_{B}&=&-\left\{2c_1\langle \chi\rangle +2c_7
\hat{\chi}_{ii}-\frac{2}{m}S_v^\mu S_v^\nu k_\mu k_\nu\right\},\label{h1}\\
\Sigma^{(b)}_{B,P}&=&iC^{(b)}_{BP}\int
\frac{d^4q}{(2\pi)^4}\left[\frac{g_A}{F_{P0}}S_v\cdot
q\right]\frac{i}{v\cdot
(k-q)+i\epsilon}\nonumber\\
&&\times\frac{i}{q^2-M_{P}^2+i\epsilon}\left[-\frac{g_A}{F_{P0}}S_v\cdot q\right]\nonumber\\
&=&-C^{(b)}_{BP}\frac{g_A^2}{(4\pi F_{P0})^2}\left\{\frac{v\cdot
k_u}{4}\right.\Bigg(\Big[3M_{P}^2-2(v\cdot
k_u)^2\Big]\Bigg[R\nonumber\\
&&+\ln\left(\frac{M_{P}^2}{\mu^2}\right)\Bigg]-2\left[M_{P}^2-(v\cdot
k_u)^2\right]\Bigg)\nonumber\\
&&+\left[M_{P}^2-(v\cdot
k_u)^2\right]^{3/2}\left.\arccos\left(-\frac{v\cdot
k_u}{M_{P}}\right)\right\},\\
\Sigma^{(c)}_{B}&=&-\frac{1}{m^2}(S_v\cdot k)(v\cdot k)(S_v\cdot
k),\label{h3}
\end{eqnarray}
which correspond to Fig. \ref{fig1} (a)-(c), respectively.
\begin{eqnarray}
\Sigma_B&=&\Sigma^{(a)}_B+\sum_{P}\Sigma^{(b)}_{B,P}+\Sigma^{(c)}_B
\end{eqnarray}
with the subscripts
$B=\Xi_{cc}^{++},\,\Xi_{cc}^{+},\,\Omega_{cc}^{+}$ and
$P=\pi^{\pm,0},\,K^{\pm,0},\,\bar{K}^{0},\,\eta$. $F_{\pi 0}$, $F_{K
0}$ and $F_{\eta 0}$ are the decay constants of $\pi$, $K$ and
$\eta$, which are $0.092$, $0.112$ and $0.110$ GeV, respectively. In
addition, the coefficients $C_{BP}^{(b)}$ are given in Table
\ref{coe}.

\begin{table*}[htbp]
\centering
\begin{tabular}{c|cccccccccccc}
\toprule[1pt]
$\left(C^{(b/d/g)}_{BP}\right)^{1/2}$&$\pi^+$&$\pi^0$&$\pi^-$&$K^+$&$K^0$&$\bar{K}^0$&$K^-$&$\eta$\\\midrule[0.5pt]
$\Xi_{cc}^{++}$&$\sqrt{2}$&1&0&$\sqrt{2}$&0&0&0&$\frac{1}{\sqrt{3}}$\\
$\Xi_{cc}^{+}$&0&-1&$\sqrt{2}$&0&$\sqrt{2}$&0&0&$\frac{1}{\sqrt{3}}$\\
$\Omega_{cc}^{+}$&0&0&0&0&0&$\sqrt{2}$&$\sqrt{2}$&$-\frac{2}{\sqrt{3}}$\\
\bottomrule[1pt]
$\left(C^{(f)}_{1BP}\right)^{1/2}$&$\pi^+$&$\pi^0$&$\pi^-$&$K^+$&$K^0$&$\bar{K}^0$&$K^-$&$\eta$\\\midrule[0.5pt]
$\Xi_{cc}^{++}$&$\sqrt{2}$&1&0&$\sqrt{2}$&0&0&0&$\frac{1}{\sqrt{3}}$\\
$\Xi_{cc}^{+}$&0&-1&$\sqrt{2}$&0&$\sqrt{2}$&0&0&$\frac{1}{\sqrt{3}}$\\
$\Omega_{cc}^{+}$&0&0&0&0&0&$\sqrt{2}$&$\sqrt{2}$&$-\frac{2}{\sqrt{3}}$\\
\bottomrule[1pt]
$C^{(f)}_{2BP}$&$\pi^+$&$\pi^0$&$\pi^-$&$K^+$&$K^0$&$\bar{K}^0$&$K^-$&$\eta$\\\midrule[0.5pt]
$\Xi_{cc}^{++}/\Xi_{cc}^{+}/\Omega_{cc}^{+}$&2&2&2&2&2&2&2&2\\
\bottomrule[1pt]
$C^{(f)}_{3BP}$&$\pi^+$&$\pi^0$&$\pi^-$&$K^+$&$K^0$&$\bar{K}^0$&$K^-$&$\eta$\\\midrule[0.5pt]
$\Xi_{cc}^{++}$&$4B_0m_d$&$2B_0(m_u)$&0&$4B_0m_s$&0&0&0&$\frac{2}{3}B_0(m_u)$\\
$\Xi_{cc}^{+}$&0&$2B_0(m_d)$&$4B_0m_u$&0&0&0&$\frac{2}{3}B_0(m_d)$\\
$\Omega_{cc}^{+}$&0&0&0&0&0&$4B_0(m_d)$&$4B_0(m_u)$&$\frac{8}{3}B_0(m_s)$\\
\bottomrule[1pt]
$C^{(f)}_{4BP}$&$\pi^+$&$\pi^0$&$\pi^-$&$K^+$&$K^0$&$\bar{K}^0$&$K^-$&$\eta$\\\midrule[0.5pt]
$\Xi_{cc}^{++}$&$4B_0m_u$&$2B_0(m_u)$&0&$4B_0m_u$&0&0&0&$\frac{2}{3}B_0(m_u)$\\
$\Xi_{cc}^{+}$&0&$2B_0(m_d)$&$4B_0m_d$&0&$4B_0(m_d)$&0&0&$\frac{2}{3}B_0(m_d)$\\
$\Omega_{cc}^{+}$&0&0&0&0&0&$4B_0(m_s)$&$4B_0(m_s)$&$\frac{8}{3}B_0(m_s)$\\
\bottomrule[1pt]
$C^{(f)}_{5BP}$&$\pi^+$&$\pi^0$&$\pi^-$&$K^+$&$K^0$&$\bar{K}^0$&$K^-$&$\eta$\\\midrule[0.5pt]
$\Xi_{cc}^{++}/\Xi_{cc}^{+}/\Omega_{cc}^{+}$&$4B_0m_u$&$2B_0(m_u+m_d)$&$4B_0m_d$&$4B_0m_u$&$4B_0m_d$&$4B_0m_s$&$4B_0m_s$&$\frac{2}{3}B_0(m_u+m_d+4m_s)$\\
\bottomrule[1pt]
\end{tabular}
\caption{The values of the coefficients
$\left(C^{(b/d/g)}_{BP}\right)^{1/2}$,
$\left(C^{(f)}_{1BP}\right)^{1/2}$, $C^{(f)}_{2BP}$,
$C^{(f)}_{3BP}$, $C^{(f)}_{4BP}$, and $C^{(f)}_{5BP}$ in Eqs.
(\ref{h1})-(\ref{h3}) and (\ref{m1})-(\ref{m2}).}\label{coe}
\end{table*}


We notice that there are three low energy constants $c_1$, $c_7$,
and $g_A$ in Eqs. (\ref{h1})-(\ref{h3}). Among these LECs, $c_1$
appearing in the next-to-leading order Lagrangian can be absorbed
into the bare mass term. Thus, $c_7$ and $g_A$ are the two unknown
constant. Due to the absence of the corresponding experimental
information, we have to fix these unknown constants based on the
other theoretical calculations. In Ref. \cite{Hu:2005gf}, Hu and
Mehen constructed Lagrangian with the following form
\begin{eqnarray}
\mathcal{L}&=&\text{Tr}[T^\dag_a(iD_0)_{ba}T_b]-g\text{Tr}[T^\dag_aT_b\vec{\sigma}\cdot
\vec{A}_{ba}]+...\label{mehen}
\end{eqnarray}
by considering the heavy diquark symmetry, where
$T_{a,i\beta}=\sqrt{2}(\Xi^*_{a,i\beta}+\frac{1}{\sqrt{3}}\Xi_{a,\gamma}\sigma^i_{\gamma
\beta})$. In Eq. (\ref{mehen}), the coupling $g=0.6$ is determined
by fitting the $D^{*+}$ width. Comparing our effective Lagrangian
with that in Eq. (\ref{mehen}), we get $g_A=g$. In the following, we
take $g_A=0.6$.

The LEC $c_7$ and bare mass $m_0$ are still unknown.
We try to fix these two unknown constants by fitting the lattice data with pion mass upto 0.4 GeV in Ref.
\cite{Alexandrou:2012xk}.

The masses of $\Xi_{cc}$ are given for different $m_\pi$ and $m_c$ in Ref. \cite{Alexandrou:2012xk}.
We assume only the bare mass ($m_0$) depends on the mass ($m_c$) of the valence charm quark,
and the dependence respects the heavy quark expansion
\begin{equation}
m_0=\tilde m_0+2 m_c+\alpha/m_c+O(1/m_c^2). \label{eqm0mc}
\end{equation}
The physical mass $m_c|_{\rm phy}$ is tuned to reproduce the mass of $D$ meson at the physical point in Ref. \cite{Alexandrou:2012xk}
\begin{equation}
m_c|_{\rm phy}=0.591\pm0.028~{\rm GeV}. \label{eqmcPhy}
\end{equation}

We give the fitted results with $\chi^2_{\rm dof}\lesssim 1$ in Table \ref{tabFit}.
Generally speaking, it indicates the lattice data are over-fitted if a result with $\chi^2_{\rm dof}\lesssim 1$ could be obtained.
One can fit the lattice data well with any $c_7$ lying in the range (-6.0, 0.6) from the table.
Therefore the current lattice data of $m_{\Xi_{cc}}$ are not enough to constrain $c_7$ yet.
However, $c_7$ being around -0.2 might be a real solution
considering that the mass of $\Omega_{cc}$ is 3.68$\sim$3.76 GeV by lattice QCD groups
 \cite{Lewis:2001iz, Na:2008hz, Liu:2009jc, Namekawa:2012mp, Alexandrou:2012xk}.
\begin{table}[htbp]
\renewcommand\arraystretch{1.5}
\centering
\caption{Parameters for fitting the lattice data from Ref.  \cite{Alexandrou:2012xk}
and the physical masses of doubly charmed baryons with the corresponding fitted parameters.
The error of the masses are from the error of $m_c|_{\rm phy}$.}\label{tabFit}
\begin{tabular}{ccc|c|cc}
\toprule[1pt]
$c_7$&$\tilde m_0$&$\alpha$&$\chi^2_{\rm dof}$&$m_{\Xi_{cc}}$&$m_{\Omega_{cc}}$\\
\hline
 0.6&    3.314&    -0.518 &   1.0&  $3.710^{+.096}_{-.100}$&  $3.045^{+.096}_{-.100}$\\
 0.3&    3.363&    -0.505 &   0.8&  $3.690^{+.095}_{-.099}$&  $3.297^{+.095}_{-.099}$\\
 0.0&    3.450&    -0.510 &   0.7&  $3.677^{+.095}_{-.099}$&  $3.557^{+.095}_{-.099}$\\
-0.1&    3.472&    -0.509 &   0.6&  $3.672^{+.095}_{-.099}$&  $3.642^{+.095}_{-.099}$\\
-0.2&    3.460&    -0.488 &   0.6&  $3.665^{+.093}_{-.097}$&  $3.726^{+.093}_{-.097}$\\
-0.3&    3.517&    -0.506 &   0.5&  $3.661^{+.095}_{-.099}$&  $3.813^{+.095}_{-.099}$\\
-0.4&    3.541&    -0.506 &   0.5&  $3.655^{+.095}_{-.099}$&  $3.898^{+.095}_{-.099}$\\
-0.5&    3.562&    -0.503 &   0.4&  $3.650^{+.095}_{-.098}$&  $3.983^{+.095}_{-.098}$\\
-1.0&    3.552&    -0.427 &   0.4&  $3.618^{+.089}_{-.092}$&  $4.405^{+.089}_{-.092}$\\
-2.0&    3.900&    -0.484 &   0.1&  $3.567^{+.093}_{-.097}$&  $5.261^{+.093}_{-.097}$\\
-4.0&    4.351&    -0.458 &   0.4&  $3.457^{+.091}_{-.095}$&  $6.966^{+.091}_{-.095}$\\
-6.0&    4.801&    -0.431 &   1.6&  $3.347^{+.089}_{-.092}$&  $8.671^{+.089}_{-.092}$\\
\bottomrule[1pt]
\end{tabular}
\end{table}

\begin{figure}[htpb]
\centering
\includegraphics[width=0.9\linewidth]{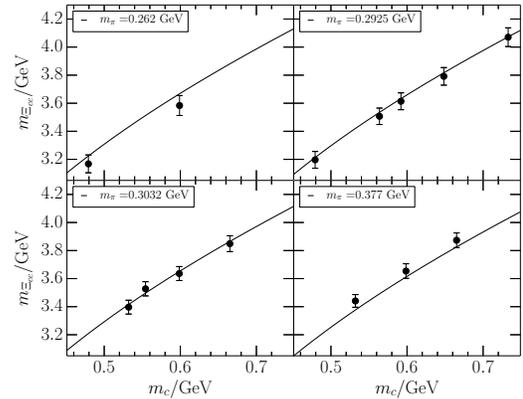}
\caption{The masses of $\Xi_{cc}$ as a function of $m_c$ for different masses of pion.
The lattice data are from Ref. \cite{Alexandrou:2012xk}, and the solid curves are our fitted results upto next to the leading order with $c_7=-0.2$, $\tilde m_0=3.460$, $\alpha=-0.488$, and $\chi^2_{\rm dof}=0.6$
\label{figFit}}
\end{figure}

We plot the best fitted results with supposing $c_7=-0.2$ in Fig. \ref{figFit}.
The best fitting needs $\tilde m_0=3.460$ and $\alpha=-0.488$ and predicts
\begin{equation}
m_{\Xi_{cc}}=3.665^{+.093}_{-.097}~{\rm GeV},\quad
m_{\Omega_{cc}}=3.726^{+.093}_{-.097}~{\rm GeV}.
\end{equation}

We have also obtained the mass correction of the doubly charmed
baryon up to the next-next-next-Leading order (N$^3$LO), which is
collected in the Appendix A. Unfortunately there appear too many
unknown LECs which cannot be fixed by experimental or theoretical
approaches. We are unable to use the N$^3$LO mass formula to compare
with the current experimental data. However, the chiral structure
and expression of the mass at the N$^3$LO will be helpful to the
chiral extrapolation of the lattice data in the lattice QCD
simulation.

\section{The chiral correction to the axial current}\label{sec4}

\begin{figure}[htbp]
\centering
\includegraphics[width=1\linewidth]{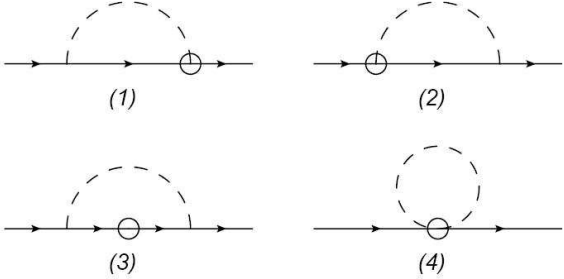}
\caption{The diagrams contributing to the renormalization of the
axial current. The circle represents an insertion of the axial
current.
\label{current}}
\end{figure}

In the following, we discuss the chiral correction to the axial
current of the doubly charmed baryon. Using Lagrangian
$\mathcal{L}^\prime_{(1)}$ and $\mathcal{L}^\prime_{(3)}$ in Eq.
(\ref{m1}), we obtain the axial current at the tree level
\begin{eqnarray}
A^{k,\mu}&=&\frac{\partial\mathcal{L}}{\partial r_k^\mu}-\frac{\partial\mathcal{L}}{\partial l_k^\mu}\nonumber\\
&=&\frac{1}{2}v^\mu \bar{H}(u^\dag T^ku-uT^ku^\dag)H\nonumber\\&&+g_A\bar{H}S_v^\mu (u^\dag T^k u+uT^ku^\dag)H\nonumber\\
&&+h_1\bar{H}S_v^\mu\langle \chi_+\rangle (u^\dag T^k u+uT^ku^\dag)H\nonumber\\&&+h_2\bar{H}S_v^\mu\{ \hat{\chi}_+, (u^\dag T^k u+uT^ku^\dag)\}H\nonumber\\
&&+h_3\bar{H}S_v^\mu\langle \hat{\chi}_+ (u^\dag T^k
u+uT^ku^\dag)\rangle H.
\end{eqnarray}
We collect the diagrams contributing to the renormalization of the
axial currents in Fig. \ref{current}.

The renormalized matrix element of $A^{k,\mu}$ between the doubly
heavy baryon states can be written as
\begin{eqnarray}
&&\langle B_d|A^{k,\mu}|B_a\rangle \nonumber\\&&=\left\{g_A2T^k_{ad}\left[1-\sum_P\frac{g_A^2}{(4\pi F_{P0})^2}\left(\frac{3M_P^2}{4}\left(R+\ln{\frac{M_P^2}{\mu^2}}\right)\right.\right.\right.\nonumber\\
&&\left.\quad+\frac{1}{2}M_P^2\right)\left(C^{(b)}_{B_aP}+C^{(b)}_{B_dP}\right)\left.\frac{1}{2}\right]+h_1\langle\chi_+\rangle 2T^k_{ad}\nonumber\\
&&\quad+h_2\sum_i\left[2\chi_{ai}T^k_{id}+2T^k_{ai}\chi_{id}\right]+h_32\langle \chi_+T^k\rangle\nonumber\\
&&\quad\left.+\sum_{P,b,c}\Sigma^{current}_{bc}\right\}\bar{u}_aS_v^\mu
u_d.\label{g1}
\end{eqnarray}
In the above equations, we have
$$\Sigma^{current}=\Sigma^{current}_{(1)}+\Sigma^{current}_{(2)}+\Sigma^{current}_{(3)}+\Sigma^{current}_{(4)},$$
where
\begin{eqnarray}
\Sigma^{current}_{(1)}=0,\label{a1}
\end{eqnarray}
\begin{eqnarray}
\Sigma^{current}_{(2)}=0,\label{a2}
\end{eqnarray}
\begin{eqnarray}
\Sigma^{current}_{(3)}&=&-\frac{g_A^3}{6F_{P0}^2}C_{\tilde{P}ab}\delta_{cd}\delta_{ab}T^k_{bc}\left\{-2(v\cdot k)\left[-4\frac{1}{32\pi^2}(R-\frac{2}{3})\right.\right.\nonumber\\
&&\times\frac{v\cdot k}{8\pi^2}\left(1-2\ln\frac{M_{\tilde{P}}}{\mu}\right)-\frac{1}{4\pi^2}\sqrt{M_{\tilde{P}}^2-(v\cdot k)^2}\nonumber\\
&&\times\left.\arccos{\frac{-v\cdot k}{M_{\tilde{P}}}}\right]+\left[M_{\tilde{P}}^2-(v\cdot k)^2\right]\left[-4\frac{1}{32\pi^2}(R-\frac{2}{3})\right.\nonumber\\
&&+\frac{1}{8\pi^2}\left(1-2\ln{\frac{M_{\tilde{P}}}{\mu}}\right)+\frac{v\cdot k}{2\pi^2}\left[M_{\tilde{P}}^2-(v\cdot k)^2\right]^{-1/2}\nonumber\\
&&\times\left.\arccos{\frac{-v\cdot k}{M_{\tilde{P}}}}-\frac{1}{4\pi^2}\right]-2M_{\tilde{P}}^2\left(\frac{1}{32\pi^2}\left(R-\frac{2}{3}\right)\right.\nonumber\\
&&\left.\left.+\frac{1}{16\pi^2}\ln{\frac{M_{\tilde{P}}}{\mu}}\right)\right\},
\end{eqnarray}
\begin{eqnarray}
\Sigma^{current}_{(4)}&=&-\frac{ig_A}{4F_{P0}^2}\left[-C_{\tilde{P}{bc}}T^k_{ab}\delta_{bd}+2C_{\tilde{P}{ab}}T^k_{bc}\delta_{ad}\delta_{bc}\right.\nonumber\\&&\left.-T^k_{cd}C_{\tilde{P}ab}\delta_{ac}\right]\frac{M_{\tilde{P}}^2}{(4\pi)^2}\left[-R+\ln\frac{\mu^2}{M_{\tilde{P}}^2}\right],
\end{eqnarray}
which correspond to Fig. \ref{current} (1)-(4), respectively. Since
the doubly heavy baryons are very heavy, we do not take into account
the small recoil corrections in the present work. The contribution
from Fig. \ref{current} (1)-(2) vanishes as shown in Eqs.
(\ref{a1})-(\ref{a2}). In the above equations, the matrices
$C_{\tilde{P}}$ ($\tilde{P}=\pi,K,\eta$) are defined as
\begin{eqnarray}
C_{\pi}=\left(
          \begin{array}{ccc}
            1 & 2 & 0 \\
            2 & 1 & 0 \\
            0 & 0 & 0 \\
          \end{array}
        \right),
C_{K}=\left(
          \begin{array}{ccc}
            0 & 0 & 2 \\
            0 & 0 & 2 \\
            2 & 2 & 0 \\
          \end{array}
        \right),
C_{\eta}=\left(
          \begin{array}{ccc}
            \frac{1}{3} & 0 & 0 \\
            0 & \frac{1}{3} & 0 \\
            0 & 0 & \frac{4}{3} \\
          \end{array}
        \right).
\end{eqnarray}
The wave function renormalization constant is expressed as
\begin{eqnarray}
Z_{N,BP}=1-C^{(b)}_{BP}\frac{g_A^2}{(4\pi F_{P0})^2}\left\{\frac{3M_P^2}{4}\left(R+\ln\frac{M_P^2}{\mu^2} \right)+\frac{1}{2}M_P^2\right\}.\nonumber\\
\end{eqnarray}
We define the axial charge of the heavy baryon through the matrix
element
\begin{eqnarray}
\langle B_d|A^{k,\mu}|B_a\rangle=g^A_{ad}\bar{u}_dS_v^\mu u_a,
\end{eqnarray}
where $g_{ad}^A$ is the axial charge.

In Eq. (\ref{g1}), there exist three low energy constants $h_1$,
$h_2$ and $h_3$. The LEC $h_1$ can be absorbed into the $g_A$ term.
There remain two unknown constants $h_2$ and $h_3$. At present,
there is no enough information to fix $h_2$ and $h_3$.  As a
crude approximation, we simply parameterize $h_2$ and $h_3$ as
$h_2=h_3= \frac{1}{\lambda^2}$, where $\lambda$ is the typical
energy scale around the mass of the doubly heavy baryons. Taking the
typical value $\lambda=\pm 3.6$ GeV, we obtain
$g^{1+i2}_{\Xi_{cc}^{++}\Xi_{cc}^+}=1.15$ and
$g^{4+i5}_{\Xi_{cc}^{++}\Omega_{cc}^+}=1.18$. If only considering
the tree level contribution, we get
$g^{1+i2}_{\Xi_{cc}^{++}\Xi_{cc}^+}=g^{4+i5}_{\Xi_{cc}^{++}\Omega_{cc}^+}=1.2$.
If $\lambda$ varies from $2$ to $5$ GeV, the range of
$g^{1+i2}_{\Xi_{cc}^{++}\Xi_{cc}^+}$ and
$g^{4+i5}_{\Xi_{cc}^{++}\Omega_{cc}^+}$ will be $1.16\sim 1.14$ and
$1.35\sim 1.14$ respectively. We also consider the case $h_{2,3}\sim -{1\over \lambda^2}$. And $g^{1+i2}_{\Xi_{cc}^{++}\Xi_{cc}^+}$ and
$g^{4+i5}_{\Xi_{cc}^{++}\Omega_{cc}^+}$ will be $1.12\sim 1.14$ and
$0.86\sim 1.07$ respectively, when $\lambda$ is in the range $2\sim 5$ GeV.

\section{Summary}\label{sec5}

Although the doubly heavy baryons have not been established
experimentally, these systems are particularly interesting. To a
large extent, they are even simpler than the light baryons such as
nucleons where the interaction among the three light quarks is very
complicated. In contrast, the presence of the two heavy quarks acts
as a static color source in the heavy quark limit. For example, the
chiral dynamics of the doubly heavy baryons is solely governed by
the light quark. We can gain valuable insights into the light quark
chiral behavior through the chiral perturbation theory study of the
doubly heavy baryons.

In this work, we have constructed the chiral effective Lagrangians
describing the interactions of light mesons and doubly charmed
baryons. We further make the non-relativistic reduction and obtain
the chiral Lagrangians up to $O(p^4)$ in the heavy baryon limit. We
have derived the chiral corrections to the mass of the doubly heavy
baryons up to N$^3$LO. Unfortunately there exist too many unknown
low energy constants. We are forced to perform the numerical
analysis at the N$^2$LO. The mass splitting of $\Xi_{cc}$ and
$\Omega_{cc}$ at the NNLO depends on one unknown low energy constant
$c_7$.
{By fitting the lattice data for $\Xi_{cc}$ from Ref. \cite{Alexandrou:2012xk} and supposing $c_7=-0.2$,
we estimate the mass of $\Omega_{cc}$ to be around 3.726 GeV,}
which may be helpful to further experimental search of the doubly charmed baryon
$\Omega_{cc}$.

Moreover, we have also performed a systematical analysis of the
chiral corrections to the axial currents and axial charges of the
doubly heavy baryons, which may be measured through the semileptonic
decays of the heavy baryons in the future.

The chiral corrections to the mass of the doubly heavy baryons have
been derived up to N$^3$LO and the axial charge to N$^2$LO. The
chiral structure and analytical expressions will be very useful to
the chiral extrapolations of the future lattice QCD simulations of
the doubly heavy baryons.

The exploration of the doubly charmed baryons is still an important
and intriguing research topic, which can deepen our understanding of
hadron spectrum and nonperturbative QCD. We are looking forward to
more developments from both experimental and theoretical studies.
There is very good chance that these doubly charmed baryons will be
observed at facilities such as LHC and BelleII.

\section*{Acknowledgement}
We would like to thank Xiu-Lei Ren, Nan Jiang and all the members of ``the workshop of chiral effective field theory-2014" that held in Sichuan University. This project is supported by the National Natural Science Foundation
of China under Grants No. 11222547, No. 11175073, No. 11035006, No.
11375240 and No. 11261130311, the Ministry of Education of China
(FANEDD under Grant No. 200924, SRFDP under Grant No. 2012021111000,
and NCET), the China Postdoctoral Science Foundation under Grant No.
2013M530461, and the Fok Ying Tung Education Foundation (Grant No.
131006). This work was also supported by the University of Adelaide and the
Australian Research Council grant FL0992247.

\section*{Appendix A: The N$^3$LO contribution to
the mass of the doubly charmed baryon}
We list the N$^3$LO chiral corrections to the mass of the doubly
charmed baryon, i.e.,
\begin{eqnarray}
\Sigma^{(d)}_{B,P}&=&C^{(d)}_{BP}i\int \frac{d^4q}{(2\pi)^4}(-)\frac{g_A}{2mF_0}S_v^\mu v^\nu (-q_\mu q_\nu+2q_\nu k_\mu)\nonumber\\
&&\times\frac{i}{q^2-M^2+i\epsilon}\frac{i}{v\cdot (k-q)+i\epsilon}(-)\frac{g_A}{F_0}S_v^\alpha q_\alpha\nonumber\\
&=&C^{(d)}_{BP}\frac{g_A^2}{2mF_0^2}(S_v\cdot
S_v)\left\{\frac{1}{4}\left(\frac{M^2}{4\pi}\right)^2\left[-R+\ln\left(\frac{\mu^2}{M^2}\right)+\frac{1}{2}\right]\right.\nonumber\\
&&\left.+v\cdot kC_{21}(v\cdot k, M^2) \right\}\label{m1}
\end{eqnarray}
from Fig. \ref{fig1} (d), where
\begin{eqnarray}
C_{21}(v\cdot k, M^2)&=&\frac{1}{n-1}\left\{(v\cdot
k)I(0)+[M^2-(v\cdot k)^2]J(0, v\cdot k)\right\},\nonumber\\
I(0)&=&\frac{M^2}{16\pi^2}\left[R+\ln\left(\frac{M^2}{\mu^2}\right)\right]+O(n-4),\nonumber\\
J(0,
\omega)&=&\frac{\omega}{8\pi^2}\left[R+\ln\left(\frac{M^2}{\mu^2}\right)-1\right]+\frac{1}{4\pi^2}\sqrt{M^2-\omega^2}\nonumber\\
&&\arccos\left(-\frac{\omega}{M}\right)+O(n-4),\nonumber\\
R&=&\frac{2}{n-4}-\left[\ln(4\pi)+\Gamma^\prime(1)+1\right].\nonumber
\end{eqnarray}
We also have the relation
\begin{eqnarray}
\Sigma^{(e)}_{B,P}&=&\Sigma^{(d)}_{B,P},
\end{eqnarray}
where $\Sigma^{(e)}_{B,P}$ comes from Fig. \ref{fig1} (e). The
corrections from Fig. \ref{fig1} (f) and $(g)$ are
\begin{eqnarray}
\Sigma^{(f)}_{B,P}&=&i\int \frac{d^4q}{(2\pi)^4} i\left\{-\frac{c_1}{F_0^2}C^{(f)}_{5BP}+\frac{1}{2}C^{(f)}_{2BP}\frac{c_2}{F_0^2}v^\mu v^\nu iq_\mu(-)iq_{\nu}\right.\nonumber\\
&&+C^{(f)}_{1BP}\left(c_3-\frac{g_A^2}{8m}\right)\frac{1}{F_0^2}v^\mu v^\nu iq_\mu (-iq_\nu)\nonumber\\
&&+\frac{1}{2}C^{(f)}_{2BP}\frac{c_4}{F_0^2}iq^\mu(-iq_\mu)+C^{(f)}_{1BP}\frac{c_5}{2F_0^2}iq^\mu (-iq_\mu)\nonumber\\
&&-\frac{c_7}{2F_0^2}C^{(f)}_{3BP}-\frac{c_7}{4F_0^2}C^{(f)}_{4BP}-\frac{c_7}{4F_0^2}C^{(f)}_{4BP}\nonumber\\
&&\left.+\frac{c_7}{3F_0^2}C^{(f)}_{5BP}\right\}\frac{i}{q^2-M^2+i\epsilon}\nonumber\\
&=&\left[\left(-\frac{c_1}{F_0^2}+\frac{c_7}{3F_0^2}\right)C^{(f)}_{5BP}-\frac{1}{2}C^{(f)}_{2BP}\frac{c_7}{2F_0^2}C^{(f)}_{4BP}+\right.\nonumber\\
&&\left.-\frac{c_7}{2F_0^2}C^{(f)}_{3BP}\right]
\left(\frac{M}{4\pi}\right)^2\left(\frac{2}{\epsilon}-r_E+1+\ln\left(\frac{4\pi\mu^2}{M^2}\right)\right.\nonumber\\
&&\left.+O(\epsilon)\right)+\frac{1}{4}\left[\frac{c_2}{2F_0^2}C^{(f)}_{2BP}+\left(c_3-\frac{g_A^2}{8m}\right)\frac{1}{F_0^2}C^{(f)}_{1BP}\right]\nonumber\\
&&\times\left(\frac{M^2}{4\pi}\right)^2\left(\frac{2}{\epsilon}-r_E+\frac{3}{2}+\ln\left(\frac{4\pi\mu^2}{M^2}\right)+O(\epsilon)\right)\nonumber\\
&&+\left[\frac{c_4}{2F_0^2}C^{(f)}_{2BP}+\frac{c_5}{2F_0^2}C^{(f)}_{1BP}\right]\left(\frac{M^2}{4\pi}\right)^2\nonumber\\
&&\times\left(\frac{2}{\epsilon}-r_E+1+\ln(\frac{4\pi\mu^2}{M^2})+O(\epsilon)\right),
\end{eqnarray}
and
\begin{eqnarray}
\Sigma^{(g)}_{B,P}&=&i\int \frac{d^4q}{(2\pi)^4}i(-)\frac{g_A}{F_0}S_v^\mu iq_\mu \frac{i}{q^2-M^2+i\epsilon}\frac{i}{v\cdot (k-q)+i\epsilon}\nonumber\\
&&\times\left[i2c_1\langle\chi\rangle+i2\chi_{jj}+i\frac{2}{m}S_v^\alpha S_v^\beta i(k-q)_\alpha i(k-q)_\beta\right]\nonumber\\
&&\times\frac{i}{v\cdot (k-q)}i(-)\frac{g_A}{F_0}S_v^\nu (-i)q_\nu C^{(g)}_{BP}\nonumber\\
&=&-\frac{1}{2}\left(c_1\langle
\chi\rangle+c_7\chi_{jj}\right)C^{(g)}_{BP}\frac{g_A^2}{F_0^2}\left\{G_2(v\cdot
k)-nG_2(v\cdot k)\right\}\nonumber\\
&&+\frac{1}{8m}C^{(g)}_{BP}\frac{g_A^2}{F_0^2}\left\{v^\mu v^\nu\Delta_{\mu\nu}-2\Delta_\mu^\mu +k^2\Delta\right.\nonumber\\
&&+(v\cdot k)^2r^2G_0(v\cdot k)+(v\cdot k)^2(nG_2(v\cdot k)+G_3(v\cdot k))\nonumber\\
&&-2(v\cdot k)^3G_1(v\cdot k)+2(v\cdot k)^2k^2J_0(v\cdot k)\nonumber\\
&&+2(v\cdot k)(nJ_2(v\cdot k)+J_3(v\cdot k))-4(v\cdot k)^2J_1(v\cdot k)\nonumber\\
&&-M^2k^2G_0(v\cdot k)-M^2(nG_2(v\cdot k)+G_3(v\cdot k))\nonumber\\
&&+2M^2(v\cdot k)G_1(v\cdot k),
\end{eqnarray}
respectively, where
\begin{eqnarray}
\Delta_{\mu\alpha}=-g_{\mu\alpha}\frac{M^2}{4}\left(\Delta-\frac{M^2}{32\pi^2}\right),
\end{eqnarray}
and $G_i$ and $J_i$ (i=0, 1, 2, 3) are defined in the appendix of Ref. \cite{Bernard:1995dp}.
\begin{eqnarray}
\Sigma^{(h)}_{B}&=&-4e_1\langle \chi\rangle\langle
\chi\rangle-e_2\left[4\chi_{i_Bi_B}\langle
\chi\rangle-\frac{4}{3}\langle
\chi\rangle\langle \chi\rangle\right]\nonumber\\
&&-e_3\left[4\langle \chi \chi\rangle-\frac{8}{3}\langle
\chi\rangle\langle \chi\rangle+\frac{4}{9}\langle \chi\rangle\langle
\chi\rangle\right]\nonumber\\
&&-e_4\left[4\sum_j\chi_{i_Bj}\chi_{ji_B}-\frac{8}{3}\chi_{i_Bi_B}\langle
\chi\rangle+\frac{4}{9}\langle \chi\rangle\langle \chi\rangle\right]\nonumber\\
&&-\frac{c_7}{m^2}S_v\cdot (ik)\left[\chi_{i_Bi_B}-\frac{1}{3}\langle
\chi\rangle\right]S_v\cdot (ik)\nonumber\\
&&-\frac{c_1}{m^2}S_v\cdot (ik)\langle \chi\rangle S_v\cdot (ik)\nonumber\\
&&+\frac{1}{2m^3}S_v\cdot (ik)[v\cdot(ik)]^2 S_v\cdot (ik)\label{m2}
\end{eqnarray}
corresponding to Fig. \ref{fig1} (h). Here $i_{\Xi_{cc}^{++}}=1$, $i_{\Xi_{cc}^+}=2$, $i_{\Omega_{cc}^+}=3$. And some coefficients in the above
expressions are listed in Table \ref{coe}.

\section{Appendix B: Some expressions}

The expressions of $\mathcal{A}_{(i)}$, $\mathcal{B}_{(j)}$ and
$\mathcal{C}_{(k)}$ that appear in Eqs. (\ref{t1})-(\ref{t3}) are
\begin{eqnarray}
\mathcal{A}_{(1)}&=&iv\cdot D+g_AS_v\cdot u,\\
\nonumber \mathcal{A}_{(2)}&=&c_1\langle
\chi_+\rangle+\frac{c_2}{2}\langle (v\cdot u)^2\rangle+c_3(v\cdot
u)^2+\frac{c_4}{2}\langle u^2\rangle+\frac{c_5}{2}u^2\\\nonumber
&&+\frac{c_6}{2}[S_v^\mu, S_v^\nu][u_\mu,
u_\nu]+c_7\hat{\chi_+}-\frac{ic_8}{4m}[S_v^\mu,
S_v^\nu]f^+_{\mu\nu}\\
&&-\frac{ic_9}{4m}[S_v^\mu, S_v^\nu]\langle
f^+_{\mu\nu}\rangle,\\
\mathcal{A}_{(3)}&=&h_1S_v^\mu \langle \chi_+ \rangle
u_\mu+h_2S_v^\mu\{\hat{\chi}_+,u_\mu\}+h_3S_v^\mu\langle
\hat{\chi}_+u_\mu\rangle+...,
\\
\nonumber \mathcal{A}_{(4)}&=&e_1\langle \chi_+\rangle\langle
\chi_+\rangle+e_2\hat{\chi}_+\langle \chi_+\rangle+e_3\langle
\hat{\chi}_+
\hat{\chi}_+\rangle+e_4\hat{\chi}_+ \hat{\chi}_+\\
&&+e_5\langle \chi_-\rangle\langle
\chi_-\rangle+e_6\hat{\chi}_-\langle \chi_-\rangle+e_7\langle
\hat{\chi}_- \hat{\chi}_-\rangle+e_8 \hat{\chi}_- \hat{\chi}_-,\\
\mathcal{B}_{(1)}&=&-2i\gamma_5S_v\cdot
D-\frac{g_A}{2}\gamma_5v\cdot u,\\\nonumber
\mathcal{B}_{(2)}&=&-\frac{c_6}{2}\gamma_5(v^\mu S_v^\nu-v^\nu
S_v^\mu)[u_\mu, u_\nu]+\frac{ic_8}{4m}\gamma_5(v^\mu S_v^\nu\\
&&-v^\nu S_v^\mu)f^+_{\mu\nu}+\frac{ic_9}{4m}\gamma_5(v^\mu
S_v^\nu-v^\nu S_v^\mu)\langle f^+_{\mu\nu}\rangle,\\
\mathcal{B}_{(3)}&=&-\frac{h_1}{2}\gamma_5\langle\chi_+\rangle v\cdot u-\frac{h_2}{2}\gamma_5\{\hat{\chi}_+,v\cdot u\}-\frac{h_3}{2}\gamma_5\langle \hat{\chi}_+v\cdot u\rangle\nonumber\\
&&+...,
\\
\mathcal{C}_{(1)}&=&iv\cdot D+2m+g_AS_v\cdot u,\\\nonumber
\mathcal{C}_{(2)}&=&-c_1\langle \chi_+\rangle-\frac{c_2}{2}\langle
(v\cdot u)^2\rangle-c_3(v\cdot u)^2-\frac{c_4}{2}\langle
u^2\rangle-\frac{c_5}{2}u^2\\\nonumber &&-\frac{c_6}{2}[S_v^\mu,
S_v^\nu][u_\mu, u_\nu]-c_7\hat{\chi_+}+\frac{ic_8}{4m}[S_v^\mu,
S_v^\nu]f^+_{\mu\nu}\\
&&+\frac{ic_9}{4m}[S_v^\mu, S_v^\nu]\langle
f^+_{\mu\nu}\rangle,\\
\mathcal{C}_{(3)}&=&h_1S_v^\mu \langle \chi_+ \rangle
u_\mu+h_2S_v^\mu\{\hat{\chi}_+,u_\mu\}+h_3S_v^\mu\langle
\hat{\chi}_+u_\mu\rangle+...,
\\
\nonumber \mathcal{C}_{(4)}&=&-e_1\langle \chi_+\rangle\langle
\chi_+\rangle-e_2\hat{\chi}_+\langle \chi_+\rangle-e_3\langle
\hat{\chi}_+
\hat{\chi}_+\rangle-e_4\hat{\chi}_+ \hat{\chi}_+\\
&&-e_5\langle \chi_-\rangle\langle
\chi_-\rangle-e_6\hat{\chi}_-\langle \chi_-\rangle-e_7\langle
\hat{\chi}_- \hat{\chi}_-\rangle-e_8 \hat{\chi}_- \hat{\chi}_-.
\end{eqnarray}

\end{document}